\documentclass[prb,twocolumn,showpacs,floats,eqsecnum,amsmath,amssymb]{revtex4}
\usepackage[dvips]{graphicx}
\def\be{\begin{equation}}
\def\ee{\end{equation}}
\def\lra{\longrightarrow}
\def\bi{\begin{itemize}}
\def\ei{\end{itemize}}
\def\bn{\begin{enumerate}}
\def\en{\end{enumerate}}
\def\bea{\begin{eqnarray}}
\def\eea{\end{eqnarray}}
\def\no{\nonumber}
\def\ba{\begin{array}}
\def\ea{\end{array}}
\def\bd{\begin{displaymath}}
\def\ed{\end{displaymath}}
\def\la{\langle}
\def\ra{\rangle}

\begin{document}



\title{Thermodynamic properties of the ferrimagnetic spin chains in the presence
of the magnetic field}

\author{J. Abouie$^1$, S. A. Ghasemi$^{1,2}$ and A. Langari$^{1,3}$}

\affiliation{
$^1$Institute for Advanced Studies in Basic Sciences,
Zanjan 45195-1159, Iran\\
$^2$Institut f\"{u}r Physik, Universit\"{a}t Basel, Basel, Switzerland\\
$^3$Physics Department , Sharif University of Technology, 
 Tehran 11365-9161, Iran
}
\date{\today}
\begin{abstract}

\leftskip 2cm \rightskip 2cm 
We have implemented three approaches
to describe the thermodynamic properties of ferrimagnetic ($S=5/2, s=2$) spin chains.
The application of cumulant expansion has been generalized to the ferrimagnetic chain in the presence of an
external magnetic field. Using cumulants, we have obtained the
field dependent effective Hamiltonian in terms of the classical
variables up to the second order of quantum corrections.
Thermodynamic functions, the internal energy, the specific heat
and the magnetic susceptibility are obtained from the effective
Hamiltonian. We have also examined the modified spin wave theory
to derive the same physical properties. Finally, 
 we have studied our model using quantum Monte Carlo
simulation to obtain  accurate results. 
The comparison of the above results and also the high
temperature series expansion shows that cumulant expansion gives
good results for moderate and high temperature regions while the
modified spin wave theory is good for low temperatures. Moreover,
the convergence regions of the cumulant expansion and the modified
spin wave theory overlap each other which propose these two  as a set of complement
methods to get the thermodynamic properties of spin models.

\end{abstract}


\maketitle

\section{Introduction}
It will become evident that there are numerous highly interesting
experimental systems which are effectively one-dimensional (1D)
models. The 1D models are more interesting from the theoretical
point of view. The quantum effects which are highlighted in the 1D
spin models represent novel physical behavior. In this class,
quantum ferrimagnets are mixed spin systems with antiferromagnetic
interactions. Mostly, they are composed of the two type of spins,
$S\neq s$. Two families of ferrimagnetic chains are described by
$ACu(pba)(H_2O)_3\cdot nH_2O$ and $ACu(pbaOH)(H_2O)_3.nH_2O$,
where $pba=1,3- propylenebis(Oxamato)$,
$pbaOH=2-hydroxo-1,3-propylenebis(Oxamato)$ and $A=Ni, Fe, Co$ and
$Mn$ \cite{verdag,blund}. Ferrimagnets, which occur rather
frequently in nature, are somehow between the antiferromagnets and
the ferromagnets. Despite the fact that the homogeneous integer
spin chains show the Haldane gap in their low energy spectrum and
the half-integer ones are gapless \cite{haldane}, 1D ferrimagnets
behave differently. The lowest energy band of the 1D ferrimagnets
is gapless which shows a ferromagnetic behavior while there is a
finite gap to the next band above it which has the
antiferromagnetic properties \cite{group-a, yamamoto1, brehmer}.
It is the acoustical and optical nature of excitations which is
the result of two different type of spins in each unit cell. This
behavior has been observed in the low and high temperature regime
of quantum ferrimagnets \cite{yamamoto1}. There are many
approaches to study the properties of the ferrimagnetic chains;
The dual features of ferrimagnetic excitations can be illuminated
by using the density-matrix renormalization group
(DMRG)\cite{group-a,sakai2} and quantum Monte Carlo (QMC)
methods\cite{brehmer, yama}. Numerical diagonalization, combined
with Lanczos algorithm\cite{abolfath} and the scaling
technique\cite{sakai1}, further have been applied to study the
modern topics such as the phase transition\cite{abolfath, yamamoto} and the quantized magnetization plateau.

The discovery of both ferromagnetic gapless and antiferromagnetic
gapped excitations have led to the investigation of the
thermodynamic properties. It has been predicted that the specific
heat at high temperatures should behave like an
antiferromagnet\cite{yamamoto1} which exhibit a Schottky peak at
the intermediate temperatures. The modified spin wave theory
(MSWT) and QMC can be used to see this behavior however, QMC is
not able to reach low temperatures sufficiently, to completely
demonstrate the ferromagnetic behavior\cite{yamamoto1}.

Most of the mentioned techniques such as QMC and DMRG have been
used for ferrimagnets with small spins. The Hilbert space of large
spins are growing exponentially and makes the computations more
difficult. In recent years there are considerable attempts which
has been focusing on the properties of new magnetic materials,
such as magnetic molecules with large effective spins ($S, s$), or
intermetallic compounds containing magnetic layers or chains.
Using MSWT\cite{yamamoto2} and high temperature series expansion
(HTSE)\cite{fukushima}, one can describe the low temperature and
high temperature properties of this systems, respectively.
Moreover, the HTSE is not enough accurate even by taking into
account higher terms (11th and 7th term for specific heat and susceptibility, respectively). 
In addition the validity regime of HTSE is
too far from the low temperature regime of MSWT to cover the full
range of temperature. The mid temperature behavior of ferrimagnets
with large values of $S$ and $s$, has not received sufficient
attention. In this respect, we used the MSWT to describe the
thermodynamic behavior of the ferrimagnetic large spin chains. We
have also implemented the QMC
simulation\cite{ghasemi} as an accurate result for comparison . At
moderate temperatures, i.e $Js<T<JSs$ ($J$ is the exchange
coupling) the results of MSWT do not coincide with QMC ones.
Therefore, to describe the physical properties at mid temperatures
we have employed the cumulant expansion (CE) \cite{fulde,kladko}.
Recently, this method has been used to study the finite
temperature behavior of large spin ferromagnetic and
antiferromagnetic systems \cite{kladko2,Garanin,abprb,abjpcm}.

In this article we have generalized the application of CE to obtain the
magnetization and magnetic susceptibility of ferrimagnets.
Moreover, the QMC simulation for ($S=5/2, s=2$) ferrimagnetic chain has been done to see the accurate behavior.
We have observed good agreement between the QMC and CE results in the intermediate and high temperature region.
The outline of this paper is as follows:
In Sec.\ref{sec2} we have employed three theoretical approaches, such as CE, QMC simulation and MSWT.
The effective Hamiltonian and magnetization in the presence of the  magnetic field are obtained using the CE up to the  second order
of $(\frac{\beta\tilde J}{s})$.
In Sec.\ref{sec3} the results and discussions have been demonstrated.
We have compared our approaches with the high temperature series expansion (HTSE).
It has been observed that CE and MSWT are two complement methods to get a good description of large spin ferrimagnetic chains
for the whole range of temperatures.

\section{theoretical approaches\label{sec2}}
\subsection{Cumulant Expansion}
Thermodynamic functions of any quantum spin system with a
Hamiltonian $\hat{H}$ can be obtained by differentiation of the
quantum partition function (${\cal Z}$) or its logarithm with
respect to the appropriate parameters. Using the basis of spin
coherent states $|\{{\bf n}_i\}\ra$ \cite{auerbach}, the trace of
an operator is reduced to the integral over a set of classical
vectors. So the partition function is reduced to that of an
effective classical spin system with the Hamilton function ${\cal
H}$ \cite{abprb, abjpcm, kladko2, Garanin}. The effective
Hamiltonian ${\cal H}$ can be expanded in terms of cumulants
\cite{fulde, kladko} of the powers of $\hat{H}$ as follows,
\bea
\no\beta {\cal H}&=&\beta\la\hat{H}\ra^c-\frac{\beta^2}{2!}\la\hat{H}\hat{H}\ra^c
+\frac{\beta^3}{3!}\la\hat{H}\hat{H}\hat{H}\ra^c+\cdots\\
&=&\beta({\cal H}^{(0)}+{\cal H}^{(1)}+{\cal H}^{(2)}+\cdots).
\label{CE:effectH1}
\eea
where $\beta=\frac{1}{T}$ and $\la O \ra^c$ represents the cumulant
of operator $O$ \cite{fulde, kladko}.
The function ${\cal H}$ evidently depends on the temperature,
thus the calculation of the physical quantities
should be done with care.

Let us consider the Hamiltonian ($\hat{H}$) of a ferrimagnetic chain which is  composed of two kinds 
of spins $S$ and $s$ ($S>s$), alternatively.
\bea
\no\hat{H}&=&\sum_{i,j}^{\frac{N}{2},\frac{N}{2}}J_{2i-1,2j}
\mathbf{S}_{2i-1}\cdot\mathbf{s}_{2j}\\
&-&\sum_{i}^{\frac{N}{2}}\mathbf{H}_{2i-1}\cdot\mathbf{S}_{2i-1}
-\sum_{i}^{\frac{N}{2}}\mathbf{H}_{2i}\cdot\mathbf{s}_{2i}.
\label{CE:Hamiltonian}
\eea
where $\mathbf{H}_i$'s are the external magnetic field at each sites.
Expressing the spin operator on each site in the coordinate
system with the $z$ axis along the coherent state vector ${\bf n}_i^z={\bf n}_i$
\bea
\no{\cal H}&=&{\cal H}^{(0)}+{\cal H}^{(1)}+{\cal H}^{(2)}\\
&&+{\cal H}^{(0)}_h+{\cal H}^{(1)}_h+{\cal H}^{(2)}_h,
\label{CE:effectH2}
\eea
where ${\cal H}^{(0)}$ is the pure classical contribution,
${\cal H}^{(1)}$ and  ${\cal H}^{(2)}$ are the quantum corrections
in the absence of a magnetic field (see Refs. [\onlinecite{abprb, abjpcm}]).
The field-dependent terms ${\cal H}_h^{(i)}$ will be expressed in the following forms,
\bea
\no{\cal H}_h^{(0)}&=&-\omega\sum_{i}^{\frac{N}{2}}\mathbf{h}_{2i-1}\cdot
\mathbf{n}_{2i-1}-\sum_{i}^{\frac{N}{2}}\mathbf{h}_{2i}\cdot\mathbf{n}_{2i},\\
\no{\cal H}_h^{(1)}&=&\frac{\beta\omega}{2s}\sum_{i,j}^{\frac{N}{2}, \frac{N}{2}}\tilde{J}_{2i-1, 2j}
\big[(\mathbf{h}_{2i-1}\cdot\mathbf{n}_{2j})\\
\no&-&(\mathbf{h}_{2i-1}\cdot\mathbf{n}_{2i-1})(\mathbf{n}_{2i-1}\cdot\mathbf{n}_{2j})
+(\mathbf{h}_{2j}\cdot\mathbf{n}_{2i-1})\\
\no&-&(\mathbf{h}_{2j}\cdot\mathbf{n}_{2j})(\mathbf{n}_{2i-1}\cdot\mathbf{n}_{2j})\big]
-\frac{\beta}{4s}\sum_i\big[\omega(1-\\
&&(\mathbf{h}_{2i-1}\cdot\mathbf{n}_{2i-1})^2)
+1-(\mathbf{h}_{2i}\cdot\mathbf{n}_{2i})^2\big],
\label{CE:effectH-h1}
\eea
where $\omega=\frac{S}{s}$.
In the above expressions $\tilde{J}=Js^2$ and ${\bf h}={\bf H}s$ are the
exchange interaction and the reduced magnetic field, respectively.
Again, ${\cal H}^{(0)}_h$ is the classical
contribution and the remaining higher orders are responsible for quantum corrections.
The field-dependent terms of order $h^3$, $Jh^2$ and $hJ^2$ in ${\cal H}^{(2)}_h$
can be calculated with the help of cumulants corresponding to the mixed field-exchange terms.
These terms are too lengthy and have been presented in the appendix.
Quasiclassical expansion up to second order of $(O(1/s^2))$ for
the internal energy and the specific heat were investigated for
different $S$, $s$ in Refs.[\onlinecite{abprb, abjpcm}].
We will now calculate the CE of the magnetization and the susceptibility.
In order to get the physical concept we have considered the nearest neighbor
interaction and that the applied fields on each site are in the same direction, i.e ${\bf h}_i=h{\bf n}$.
Thus, the field-dependent terms of the effective Hamiltonian is reduced
to the following forms:
\bea
\no{\cal H}_h^{(0)}&=&-h{\mathbf n}\cdot\sum_{i=1}^{\frac{N}{2}}({\mathbf n}_{2i}+\omega{\mathbf n}_{2i-1}),\\
\no{\cal H}_h^{(1)}&=&\frac{\beta h\omega\tilde{J}}{2s}\sum_{i=1}^N\big\{2{\mathbf n}\cdot{\mathbf n}_i-
{\mathbf n}\cdot{\mathbf n}_i({\mathbf n}_{i-1}\cdot{\mathbf n}_i+{\mathbf n}_i\cdot{\mathbf n}_{i+1})\big\}\\
\no&&-\frac{\beta h^2}{4s}\sum_{i=1}^{\frac{N}{2}}\big\{\omega(1-({\mathbf n}\cdot{\mathbf n}_{2i-1})^2)+
1-({\mathbf n}\cdot{\mathbf n}_{2i})^2\big\},\\
\no{\cal H}_{h}^{(2)}&=&\frac{-\beta^2h^3}{12s^2}\Psi_1+\frac{\beta^2\omega h^2\tilde{J}}{4s^2}\Psi_2
-\frac{\beta^2\omega h\tilde{J}^2}{4s^2}\Psi_3\\
&&-\frac{\beta^2\omega^2\tilde{J}^2h}{4s^2}\Psi_4-\frac{\beta^2\omega h\tilde{J}^2}{24s^3}\Psi_5.
\label{CE:effectH-h2}
\eea
where $\Psi_i$ is expressed in terms of the classical vectors
${\bf n}$ and ${\bf n}_i$ (see appendix).
The partition function is represented by
the effective Hamiltonian defined in the previous equations,
\bea
\no{\cal Z}&=&\big(\frac{2\omega s+1}{4\pi}\big)^{\frac{N}{2}}\big(\frac{2s+1}{4\pi}\big)^{\frac{N}{2}}
\int\Pi_{i=1}^Nd{\bf n}_ie^{-\beta{\cal H}^{(0)}}\times\\
\no&&\big[(1-\beta{\cal H}^{(0)}_h+\frac{\beta^2}{2}[{\cal H}^{(0)}_h]^2+\dots)\times\\
\no&&\big(1-\beta{\cal H}_h^{(1)}+
\frac{\beta^2}{2}[{\cal H}_h^{(1)}]^2-\beta{\cal H}^{(1)}+\beta^2{\cal H}_h^{(1)}{\cal H}^{(1)}\\
&&+\frac{\beta^2}{2}[{\cal H}^{(1)}]^2-\beta({\cal H}^{(2)}+{\cal
H}_h^{(2)})+O(1/s^3)\big) \big].
\label{CE:partition2} 
\eea 
The reduced magnetic susceptibility is obtained by two times
differentiating from the logarithm of the partition function with
respect to $h$, i.e, 
\be 
\chi=\lim_{H\rightarrow0}\frac{\partial m}{\partial h}. 
\label{CE:Suscep} 
\ee 
where $H$ is the magnetic field and $m$ is the scaled magnetization per site and given by 
\be
m=\frac{1}{N}\frac{\partial ln{\cal Z}}{\partial(\beta h)}.
\label{CE:mag1} 
\ee 
In the limit of $H\rightarrow0$, all of the
terms containing $h^3$ or higher orders of $h$ will vanish in the
partition function. So, we will keep the expansion up to the
second order of $h$ in the partition function Eq.(\ref{CE:partition2}). By the integration on the coherent states,
we find the scaled magnetization in terms of the coupling and the
magnetic field. The reduced magnetic susceptibility is as follow:
\bea
\no\chi&=&\frac{2\beta\omega}{3}\left(\frac{B}{1-B^2}\right)+\frac{\beta(1+\omega^2)}{6}\left(\frac{1+B^2}{1-B^2}\right)\\
\no&+&\frac{\beta (\omega+1)}{6s}-\frac{2\beta(1+\omega)}{3s}\left(\frac{B}{1-B}\right)\\
&-&\frac{\beta^2\tilde{J}\omega}{3s^2}\left(1-B-\frac{B}{\omega\xi}\right),
\label{CE:susc1}
\eea
where $\xi=\beta\tilde{J}$ and $B=\coth(\omega\xi)-\frac{1}{\omega\xi}$ is the Langevin function.

We have plotted in Fig.(\ref{suscep.}) the actual magnetic susceptibility
($\chi_a$) of the $(S=5/2, s=2)$
ferrimagnetic chain versus temperature.
It is related to the susceptibility defined in Eq.(\ref{CE:Suscep}) by the
following relation,
\be
\chi_a=\lim_{H\rightarrow0}\frac{\partial m_a}{\partial H}=s^2\chi,
\label{actual-susc}
\ee
where $m_a=sm$ is the actual magnetization.
The presented CE result contains the
quantum corrections up to the second order of $(\frac{\beta{\tilde J}}{s})$.
We will discuss on the quality of our results in comparison with the other ones
in the next section.
\subsection{Quantum Monte Carlo simulation}
We have implemented the quantum Monte Carlo (QMC) simulation for
the ferrimagnetic $(S=5/2, s=2)$ chain of length $N=64$. The fairly
large attainable length size gives us the thermodynamic properties
as a very good approximate of the infinite size chain. In doing
so, we have considered the Hamiltonian of
Eq.(\ref{CE:Hamiltonian}) for $N=64$ and without the magnetic field.
We utilized the QMC algorithm based on the Suzuki-Trotter
decomposition \cite{Suzuki} of the Checkerboard-Type
\cite{Hirsch}. In this respect, we begin by breaking the
Hamiltonian into four pieces, ${\hat H}=\frac{{\hat H}_0}{2}+{\hat
H}_a+\frac{{\hat H}_0}{2}+{\hat H}_b$, where ${\hat H}_0$ contains
the interactions in $z$ direction. ${\hat H}_a$ and ${\hat H}_b$
represent the interaction in the transverse direction
alternatively. The partition function is expressed by the
Suzuki-Trotter formula as the following,
\bea
\no{\cal Z}&=&Tr e^{-\beta{\hat H}}=\lim_{m\rightarrow\infty}{\cal Z}_m,\\
{\cal Z}_m:&=&Tr\left(e^{-\frac{\beta\hat{H}_0}{2m}}
e^{-\frac{\beta\hat{H}_a}{m}}e^{-\frac{\beta \hat{H}_0}{2m}}
e^{-\frac{\beta \hat{H}_b}{m}} \right)^m, \label{QMC:partition}
\eea
where $m$ is Trotter number. Performing the trace operation,
we have a two-dimensional classical Hamiltonian rather than the
one-dimensional quantum Hamiltonian. This classical Hamiltonian
has $2mN$ spins. We have considered the plaquette flip for the
evolution of the Monte Carlo simulation. The reason is related to
the huge number of single spin flips which are not permitted
because their Boltzman weight is zero. For instance, in the case
of a plaquette of four spins which contains two $S=\frac{5}{2}$
and two $s=2$, there exists 900 different configurations. There
are only 110 configurations with nonzero Boltzman weights. All of
these nonzero cases can be obtained by a plaquette flip
\cite{ghasemi}. The quantities like internal energy, heat
capacity, and magnetic susceptibility
 depend on the Trotter number ($m$). In the limit of $m\lra\infty$,
these quantities tend to their correct values. Therefore, It
should be taken the biggest possible value of $m$, especially at
low temperatures. However, when temperature decreases the
convergence relative to $m$ becomes small. In other words, using
the big value for $m$ makes two kinds of problems. Firstly,
$\frac{\beta}{m}$ becomes small so the state of the system (spin
configuration) changes hardly (evolves slowly). Secondly, when $m$
is large the global flips to change the total magnetization are
hard to be accepted at low temperatures. Consequently, many Monte
Carlo steps are needed to equilibrate the system at a large $m$
and the low temperature. Therefore, for the mentioned reasons we
have  performed the calculations for each temperature with
different values of $m$ and utilizing the least-square
extrapolation method (Eq.(\ref{am}) to find the limit of
$m\rightarrow \infty$)
\bea
A(m) = A_{\infty} + \frac{A_1}{m^2}
+\frac{A_2}{m^4} + \cdots .
\label{am}
\eea
For moderate and low
temperature regime ($T< 3J$) we have considered three different
values for the Trotter number, $m=15, 20, 30$. At higher
temperatures, the convergence happens for the lower $m$ values. To
equlibriate the system we have spent $10^5$ Monte Carlo steps and
$10^6$ steps for measurement. Accuracy of the measured quantities
depends on the temperature, for higher temperatures we got higher
accuracy. However, the error bar is less than the symbol sizes in
our plots.

The internal energy, specific heat and magnetic susceptibility of the
ferrimagnetic $(S=5/2, s=2)$
chain have been plotted in Figs. (\ref{energy}), (\ref{s-heat}) and (\ref{suscep.}) respectively.
We will discuss our results in the next section when we compare it
with the other ones.

\subsection{Modified Spin Wave Theory}
In the modified spin wave theory, usually it is considered a single-component
bosonic representation of each spin variable
at the cost of the rotational symmetry.
To simplify in the incoming calculations, we consider the following form of the Hamiltonian
for the ferrimagnetic chain:
\be
\hat{H}=J\sum_{i=1}^N({\mathbf S}_i\cdot{\mathbf s}_{i-1}+{\mathbf s}_i\cdot{\mathbf S}_{i}).
\label{SW:Hamiltonian}
\ee
Using the Holestain-Primakoff and the Bogoliubov
transformation, the Hamiltonian (\ref{SW:Hamiltonian}) is diagonalized
${\cal H}=-2NJSs+E_1+E_0+{\cal H}_1+{\cal H}_0+O(S^{-1})$,
where $E_i$ gives the $O(S^i)$ quantum corrections to the ground
state energy and ${\cal H}_i$ is expressed in terms of $\alpha_k^{\dagger}$ and
$\beta_k^{\dagger}$ and gives the quantum corrections to the dispersion relation. (see Ref.[\onlinecite{yamamoto2}]).
$\alpha_k^{\dagger}$ and $\beta_k^{\dagger}$ are the creation operators of the
ferromagnetic and antiferromagnetic spin waves with momentum $k$, respectively.

\begin{figure}
\centerline{\includegraphics[width=7cm,angle=0]{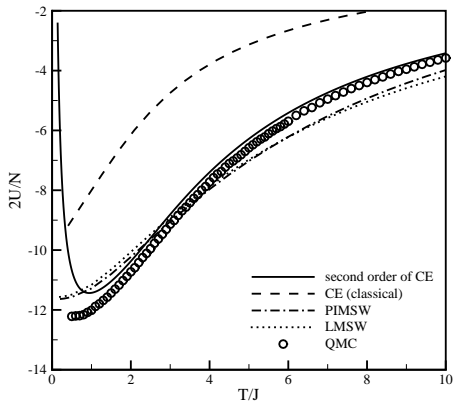}}
\caption{Temperature dependence of the internal energy per cell of the
ferrimagnetic ($S=5/2, s=2$) chain.
The solid line: Cumulant expansion up to the second order,
Dashed line: Classical part of the cumulant expansion,
DashDot line: Perturbational interacting-modified-spin-wave-theory (PIMSWT),
Doted line: Linear-modified-spin-wave-theory (LMSWT) and
circle: Quantum Monte Carlo simulation
with $N=64$ spins.}
\label{energy}
\end{figure}

At finite temperatures, we assume that
$\widetilde n^\pm_k\equiv\sum_{n^-,n^+}n^\pm P_k(n^-,n^+)$,
for the spin wave distribution functions, where
$P_k(n^-,n^+)$ is the probability of $n^-$ ferromagnetic and
$n^+$ antiferromagnetic spin waves appearing in the $k$-momentum
state and
satisfies $\sum_{n^-,n^+} P_k(n^-,n^+)=1$ for all $k$'s \cite{yamamoto2}.
The substitutions $\widetilde n^{-}_k=\alpha_k^\dagger\alpha_k$ and
$\widetilde n^{+}_k=\beta_k^\dagger\beta_k$ in the spin-wave Hamiltonian gives the
zero field free energy,
\bea
\no F&=&E_{\rm g}+\sum_k(\widetilde n^-_k\omega_k^-+\widetilde n^+_k\omega_k^+)\\
&+&T\sum_k\sum_{n^-,n^+}P_k(n^-,n^+){\rm ln}P_k(n^-,n^+)\,.
\label{E:F}
\eea
To keep the number of bosons finite, one should apply the
following constraint,
\be
\la:S^z-s^z:\ra=(S+s)N-(S+s)\sum_k\sum_{\sigma=\pm}
\frac{\widetilde n^\sigma_k}{\omega_k}=0,
\label{E:constSM}
\ee
where $\omega_k=((S-s)^2+4Ss\sin^2k)^{1/2}$ and the normal
ordering is taken with respect to $\alpha$ and $\beta$. By
minimization of the free energy (\ref{E:F}) with respect to
$P_k(n^-,n^+)$'s under the condition (\ref{E:constSM}) we can
obtain the free energy and the magnetic susceptibility at thermal
equilibrium as follows
\bea
&&F=E_{\rm
g}+\mu(S+s)N-T\sum_{k}\sum_{\sigma=\pm}{\rm ln}(1+\widetilde
n^\sigma_k),~~~~~
\label{E:MSW2SC-F}\\
&&\chi=\frac{1}{3T}\sum_k\sum_{\sigma=\pm}\widetilde
n^\sigma_k(1+\widetilde n^\sigma_k), \label{E:MSW2SC-S}
\eea
where
$\widetilde n^\pm_k=[{\rm e}^{\left(J\omega_k^\pm-\mu
(S+s)/\omega_k\right)/T}-1]^{-1}$, and $\mu$ is the Lagrange
multiplier to consider the constraint (\ref{E:constSM}). 
This set of equations has no closed analytic solution. In the case of $(S=5/2, s=2)$
we have numerically solved Eqs. (\ref{E:constSM}) and
(\ref{E:MSW2SC-F}) in the thermodynamic limit, and visualized them
in Figs. (\ref{energy}), (\ref{s-heat}) and (\ref{suscep.}).
In previous equations we have chosen $k_B=1$. $\omega_k^-$ and $\omega_k^+$ are the
ferromagnetic and antiferromagnetic excitation gaps, respectively. They have different 
values in the linear modified spin wave theory (LMSW)
 and perturbational interacting modified spin wave theory (PIMSW). 
In the  PIMSW, the $O(S^0)$ terms have been considered. 
Because the antiferromagnetic excitation gap is significantly improved by the inclusion of the $O(S^0)$
correlation, the location of the Schottky peak can be also reproduced very well by the perturbational interacting modified spin waves.
%
%
\section{Results and disscutions \label{sec3}}
We have obtained the effective Hamiltonian of the ferrimagnetic chains in the presence of
an external magnetic field to second order of cumulant expansion, Eq.(\ref{CE:effectH-h2}).
The  zeroth order term shows the classical contribution which simply represents the coupling energy of
the classical spins with the external magnetic field. Quantum corrections have a non-Heisenberg form
and they are important in the intermediate temperatures.

In Fig.(\ref{energy}), we have shown the internal energy per unit
cell of spins $(\frac{2U}{N})$ versus temperature. The big
difference between the zeroth order (classical contribution shown
by dashed line) and the second order cumulant expansion (quantum
corrections shown by solid line) shows the importance of the
corrections in the intermediate and higher temperatures. The
discrepancy is high, even in the present case of fairly large spins
($S=5/2, s=2$) which seems to behave classically. The reason is
related to the dual features of ferrimagnets, i.e the low
temperature behavior is like ferromagnets and in the high
temperatures they show antiferromagnetic behavior. There is a
spectral gap in the subspace $S_{tot}=\frac{N}{2}(S-s)+1$, where
the optical magnons play an important role. In the case of
$(S=5/2, s=2)$, the spectral gap of optical magnons at $k=0$ is;
$\Delta_0=1.36847 J$ (Ref.[\onlinecite{sakai3}]). This means that the model does not behave
pure classically. So, to describe the finite temperature behavior
of the system, we should consider the quantum corrections to the
classical part. At low temperature the second order of CE has a
large deviation in comparison with the other results. Because in
the low temperature region classical fluctuations are not strong
enough to suppress the quantum ones. However, the low temperature
region has been excluded from the convergence domain by
construction when CE is expressed as a series in the order of
$\beta J s < 1$. Obviously, the classical term is dominant at very
high temperatures.

\begin{figure}
\centerline{\includegraphics[width=7cm,angle=0]{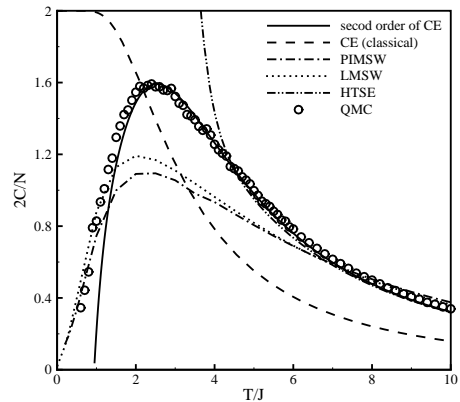}}
\caption{The specific heat per cell of the ferrimagnetic ($S=5/2, s=2$) chain.
Cumulant expansion up to second order (solid line),
classical part of the cumulant expansion (dashed line),
perturbational interacting-modified-spin-wave-theory (PIMSWT- dash-dot line),
linear-modified-spin-wave-theory (LMSWT- doted line),
high temperature series expansion (dash-dot-dot line) and
quantum Monte Carlo simulation (open circle)
with $N=64$ spins.}
\label{s-heat}
\end{figure}

\begin{figure}
\centerline{\includegraphics[width=7cm,angle=0]{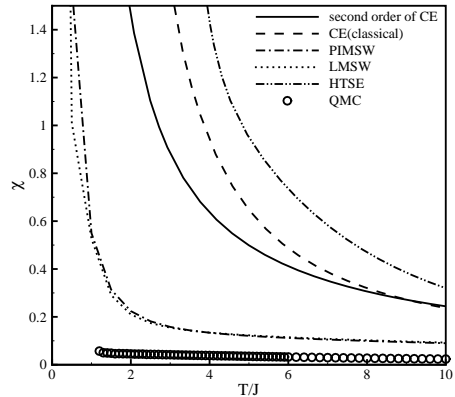}}
\caption{Magnetic susceptibility per spin of the ferrimagnetic ($S=5/2, s=2$) chain
versus temperature.
The solid line: Cumulant expansion up to second order,
Dashed line: Classical part of the cumulant expansion,
DashDot line: Perturbational interacting-modified-spin-wave-theory (PIMSWT),
Doted line: Linear-modified-spin-wave-theory (LMSWT),
DashDotDot: High temperature series expansion and
circle: Quantum Monte Carlo simulation
with $N=64$ spins.}
\label{suscep.}
\end{figure}

To have an impression on the accuracy of our results, we have plotted the results of
the QMC simulation for comparison. The second order CE of the internal energy in Fig.(\ref{energy})
fits very well on the QMC results for $T>2J$. This is actually the validity regime of our CE approach,
$T>Js$. We have also plotted in Fig.(\ref{energy}) the results of two different modified spin wave
theory which deviate slightly from the QMC ones . However, the accuracy of the different schemes
can be best visualized in the physical quantities like the specific heat and the magnetic susceptibility
which is shown in Figs.(\ref{s-heat}, \ref{suscep.}).

In Fig. (\ref{s-heat}), we have plotted the specific heat per unit cell of spins $(\frac{2C}{N})$ for a $(S=5/2, s=2)$
ferrimagnetic chain. The results of CE have been shown as the pure classical contribution
and also the whole contribution to the second order. The big difference between them
verifies the significant corrections of the second order CE.
We have also plotted the result of QMC simulation, for comparison.
We observe very good agreement between the CE and the QMC results.
According to the results presented in Figs.(\ref{energy}, \ref{s-heat}),
the QMC simulation results confirm that  the CE is a very good analytical approach
to describe the thermodynamic properties of a ferrimagnetic system
with large spins at moderate and high temperatures.

We have also shown in Fig.(\ref{s-heat}) the results of the MSWT
for the specific heat. We have examined the LMSW and the PIMSW for our system. As it is observed from Fig.
(\ref{s-heat}), both of LMSW and PIMSW can reproduce the high
temperature behavior of heat capacity close to the QMC simulation
results. Furthermore, they can show the Schottky peak at mid
temperatures. Although the PIMSW can reproduce the location of the
Schottky peak fairly well, it can not estimate the peak value well for
large spins in comparison with the CE.
The reason of this
discrepancy in the MSWT is as follows. Let us come back to the
spin wave theory and draw your attention to the bosonic
Hamiltonian; ${\cal H}=-2NJSs+E_1+E_0+{\cal H}_1+{\cal
H}_0+O(S^{-1})$ where
\bea
\no{\cal H}_i&=&J\sum_k[\omega_i^-(k)\alpha_k^{\dagger}\alpha_k+\omega_i^+(k)\beta_k^{\dagger}\beta_k\\
&&+\gamma_i(k)(\alpha_k\beta_k+\alpha_k^{\dagger}\beta_k^{\dagger})],
\eea 
and $\gamma_i(k)$'s have been introduced in
Ref.[\onlinecite{yamamoto2}]. The last two terms in ${\cal H}_i$
are the normal-ordered quassiparticle interactions. In MSWT,
whether LMSW or PIMSW we have eliminated these interactions, i.e
we choose 
\bea
\no\gamma_1(k)=0~~~\lra~~~\tanh{2\theta_k}=\frac{2\sqrt{Ss}\cos{k}}{S+s}.
\eea 
However, in the low and moderate temperatures, the
magnon-magnon interactions play an important role. Therefore it is
surmised that if we consider at least the first order of the
quasiparticle interaction (i.e $\gamma_1(k)\neq 0$ and
$\gamma_0(k)=0$ ), we can produce the Schottky peak value  more
precisely. Although the agreement between MSWT (without the
quassiparticle interaction) and the other results is not perfect,
it is a remarkable success for the spin wave theory in the
one-dimensional large spin ferrimagnet that all relevant features
are quantitatively rather well reproduced over a very large
temperature range.

Finally, we have  plotted in Fig.(\ref{suscep.}) the magnetic
susceptibility per total number of spins ($\frac{\chi}{N}$) of the
$(S=5/2, s=2)$ ferrimagnetic chain versus temperature. The result
of the second order CE is shown by the solid line. The CE result
shows qualitatively the features of the ferrimagnetic chains, i.e
the quasiclassical $\chi$ shows the divergence for $T\lra 0$ like
a ferromagnet and a Curie law ($\frac{1}{T}$) decay at high
temperatures. Meanwhile in Fig. (\ref{suscep.}) we have plotted
the result of the QMC simulation. Our simulation result shows very
well the antiferromagnetic feature of the ferrimagnetic system.
But it can not produce the ferromagnetic behavior at low
temperatures. The reason is as follow. The antiferromagnetic trait
of the model does not depend on the size of system, i.e all
antiferromagnetic features can be reproduced in a rather short
system size, while the ferromagnetic ones completely depend on the
number of spins. This means that the ferromagnetic features emerge
only slowly with the growing of system size. In the  QMC approach
we have considered 32 cells (64 spins) for simulation. The
computation time grows exponentially by going to larger sizes.
Specially, it happens in the calculation of magnetic
susceptibility to reach the equilibrium condition. However, our
main interest in this study is  the results for intermediate and
large temperature regions where reasonable values exist.

We have also shown in Figs. (\ref{s-heat}) and (\ref{suscep.}) the results of HTSE for
the specific heat and the susceptibility of $(S=5/2, s=2)$ ferrimagnetic chain, respectively.
The HTSE is an expansion in powers of $\beta J$. Recently, Fukushima and his collaborators
have implemented a suitable Pad\'{e} approximation to obtain the thermodynamic functions of
the mixed spin chains.
They have found the specific heat and the susceptibility up to
$(O(\beta J)^{11})$ and $(O(\beta )^7)$, respectivly\cite{fukushima}.
The HTSE results for the specific heat shown in Fig.(\ref{s-heat}) converges to
the QMC results at $T>4J$. However, the deviation from the QMC result is more pronounced for the
magnetic susceptibility shown in Fig.(\ref{suscep.}).

It is worth to mention the
two differences between the CE and HTSE results. Firstly, the convergence region
of CE is larger than the HTSE one, ie, the CE is valid for $T>Js$ whereas the validity of the HTSE
is for  $T >\sqrt{S(S+1)}J$. Secondly, the HTSE fails to produce the Schottky peak of the specific heat
While the CE can generate it as well as the QMC simulation.

Our results state that the combined methods of the cumulant expansion for   $T>Js$
and the modified spin wave theory for $T<Js$, give a good approximation for  the 
whole finite temperature behavior of quantum ferrimagnets.
The reason is related to the overlap of the convergence
regions of the mentioned method.


\acknowledgments

A. L. would like to acknowledge the hospitality of the
Max-Planck-Institut
f\"ur Chemische Physik fester Stoffe in Dresden where the final preparation
of this work has been done.


\appendix
\section{}
The expressions of the second term (${\cal H}_h^{(2)}$) of the
field dependent effective Hamiltonian are obtained as follows,
\bea
\no\Psi_1&=&\sum_{i=1}^{\frac{N}{2}}\{\omega{\mathbf n}\cdot{\mathbf n}_{2i-1}(1-({\mathbf n}\cdot{\mathbf n}_{2i-1})^2)\\
\no&&+{\mathbf n}\cdot{\mathbf n}_{2i}(1-({\mathbf n}\cdot{\mathbf n}_{2i})^2)\},\\
\no\Psi_2&=&\sum_{i=1}^{N}\big\{1-2({\mathbf n}\cdot{\mathbf n}_{i})^2+
  ({\mathbf n}\cdot{\mathbf n}_i)({\mathbf n}_i\cdot{\mathbf n}_{i+1})
  ({\mathbf n}_{i+1}\cdot{\mathbf n})\\
\no&-&{\mathbf n}_i\cdot{\mathbf n}_{i+1}-({\mathbf n}\cdot{\mathbf n}_i)
  ({\mathbf n}\cdot{\mathbf n}_{i+1})+
  ({\mathbf n}_i\cdot{\mathbf n}_{i+1})\times\\
\no&&\big(({\mathbf n}\cdot{\mathbf n}_{i})^2+
({\mathbf n}\cdot{\mathbf n}_{i+1})^2\big)\big\},\\
\no\Psi_3&=&\sum_{i=1}^N({\mathbf n}\cdot{\mathbf n}_{i})
   \big\{({\mathbf n}_{i-1}\cdot{\mathbf n}_{i})^2
   +({\mathbf n}_{i}\cdot{\mathbf n}_{i+1})^2\big\}\\
\no&+&\sum_{i=1}^{\frac{N}{2}}\big\{2({\mathbf n}\cdot{\mathbf n}_{2i-1})
   ({\mathbf n}_{2i-1}\cdot{\mathbf n}_{2i})({\mathbf n}_{2i-1}\cdot{\mathbf n}_{2i-2})\\
\no&+&({\mathbf n}\cdot{\mathbf n}_{2i})({\mathbf n}_{2i}\cdot{\mathbf n}_{2i-1})
({\mathbf n}_{2i-1}\cdot{\mathbf n}_{2i-2})\\
\no&+&({\mathbf n}\cdot{\mathbf n}_{2i-2})({\mathbf n}_{2i-2}\cdot{\mathbf n}_{2i-1})
   ({\mathbf n}_{2i-1}\cdot{\mathbf n}_{2i})\\
\no&-&({\mathbf n}\cdot{\mathbf n}_{2i-2}+{\mathbf n}\cdot{\mathbf n}_{2i-1}
  +{\mathbf n}\cdot{\mathbf n}_{2i})\times\\
\no&+&({\mathbf n}\cdot{\mathbf n}_{2i-2})({\mathbf n}_{2i-2}\cdot{\mathbf n}_{2i-1})
   ({\mathbf n}_{2i-1}\cdot{\mathbf n}_{2i})\\
\no&-&({\mathbf n}\cdot{\mathbf n}_{2i-2}+{\mathbf n}\cdot{\mathbf n}_{2i-1}
  +{\mathbf n}\cdot{\mathbf n}_{2i})\times\\
\no&&({\mathbf n}_{2i-1}\cdot{\mathbf n}_{2i}+{\mathbf n}_{2i-2}\cdot{\mathbf n}_{2i-1})\\
\no&-&({\mathbf n}_{2i}\cdot{\mathbf n}_{2i-2})({\mathbf n}\cdot{\mathbf n}_{2i}
  +{\mathbf n}\cdot{\mathbf n}_{2i-2})\big\}
\eea
\bea
\no\Psi_4&=&\sum_{i=1}^N({\mathbf n}\cdot{\mathbf n}_{i})
   \big(({\mathbf n}_{i-1}\cdot{\mathbf n}_{i})^2
   +({\mathbf n}_i\cdot{\mathbf n}_{i+1})^2\big)\\
\no&+&\sum_{i=1}^{\frac{N}{2}}\big\{2({\mathbf n}\cdot{\mathbf n}_{2i})
   ({\mathbf n}_{2i}\cdot{\mathbf n}_{2i-1})({\mathbf n}_{2i}\cdot{\mathbf n}_{2i+1})\\
\no&+&({\mathbf n}\cdot{\mathbf n}_{2i-1})({\mathbf n}_{2i-1}\cdot{\mathbf n}_{2i})
    ({\mathbf n}_{2i}\cdot{\mathbf n}_{2i+1})\\
\no&+&({\mathbf n}\cdot{\mathbf n}_{2i+1})({\mathbf n}_{2i+1}\cdot{\mathbf n}_{2i})
   ({\mathbf n}_{2i-1}\cdot{\mathbf n}_{2i})\\
\no&-&({\mathbf n}\cdot{\mathbf n}_{2i-1}+{\mathbf n}\cdot{\mathbf n}_{2i+1}
  +{\mathbf n}\cdot{\mathbf n}_{2i})\times\\
\no&&({\mathbf n}_{2i-1}\cdot{\mathbf n}_{2i}+{\mathbf n}_{2i}\cdot{\mathbf n}_{2i+1})\\
\no&-&({\mathbf n}_{2i-1}\cdot{\mathbf n}_{2i+1})
   ({\mathbf n}\cdot{\mathbf n}_{2i-1}+{\mathbf n}\cdot{\mathbf n}_{2i+1})\big\}\\
\no\Psi_5&=&\sum_{i=1}^{\frac{N}{2}}\big[
(1-{\mathbf n}_{2i-1}\cdot{\mathbf n}_{2i})(2{\mathbf n}\cdot{\mathbf n}_{2i}+
 2{\mathbf n}\cdot{\mathbf n}_{2i-1}\\
\no&+&(1-3{\mathbf n}_{2i-1}\cdot{\mathbf n}_{2i})
 ({\mathbf n}\cdot{\mathbf n}_{2i-1}+{\mathbf n}\cdot{\mathbf n}_{2i}))\\
\no&+&(1-{\mathbf n}_{2i-1}\cdot{\mathbf n}_{2i-2})(2{\mathbf n}\cdot{\mathbf n}_{2i-2}
+2{\mathbf n}\cdot{\mathbf n}_{2i-1}\\
\no&+&(1-3{\mathbf n}_{2i-1}\cdot{\mathbf n}_{2i-2})
({\mathbf n}\cdot{\mathbf n}_{2i-1}+{\mathbf n}\cdot{\mathbf n}_{2i-2}))
\big]
\eea



\end{document}